**Abdus Salam: A Reappraisal**

**PART I**

**How to Win the Nobel Prize**

Norman Dombey[*]

Physics and Astronomy Department

University of Sussex

Brighton BN1 9QH

September 24 2011

**ABSTRACT**

Abdus Salam's correspondence  during his time as Director of the International Centre for Theoretical Physics (ICTP)  is held in the Abdus Salam Archive of the Salam International Centre for Theoretical Physics[+].  I use  this correspondence to discuss  his contribution to the theory of electromagnetic and weak interactions for which he was awarded the Nobel Prize in Physics in 1979.

_Note on References to the Salam Archive of the Salam ICTP_

_Where possible the date of a letter or memo and the file number are both included in the reference. Unfortunately this is not always possible.  AS is short for Abdus Salam_

[*]Email n.dombey@sussex.ac.uk

[+]The ICTP was renamed the Salam International Centre for Theoretical Physics in 1997

On 15 October 1979 the Nobel Prize for physics was awarded to the Pakistani physicist Abdus Salam, then Director of the International Centre for Theoretical Physics (ICTP) at Trieste, together with two Harvard physicists, Steven Weinberg and Sheldon Glashow. This was the first time that a Moslem had won a Nobel Prize for Science. I knew Salam quite well from his time as Professor of Theoretical Physics at Imperial College, London, where I had spent the year 1963-64 before going to Sussex as a Lecturer. When I applied for the position I asked Salam for a reference. 'That's fine' he replied. 'You write it and I'll sign it. You know more about your work than I do'. That's how I got my first job. Salam was not bound by established convention.

The New York Times explained[1] that the 1979 Nobel award gave increasingly strong support to a "theory that many view as among the most momentous of this century namely that two of the basic forces in nature- -electromagnetism which turns electric motors and the 'weak' force which causes radioactive (beta-) decay in some atomic nuclei- -are facets of the same phenomenon.....The unification hypothesis was offered in 1967 by Dr Abdus Salam and Dr Steven Weinberg. ...Some aspects of it had been anticipated by Glashow in 1961". An experiment [2] earlier that year had shown that there was a slight difference in the number of right-handed and left-handed electrons scattered by hydrogen nuclei. The New York Times then commented [3] "just such a violation had been exactly predicted in the theory "propounded a decade ago by an American Steven Weinberg of Harvard University and a Pakistani, Abdus Salam of Imperial College London" .

A year before the award I had written an article[4] with my colleague David Bailin for the journal NATURE setting forward an alternative theory of the same phenomena. I received a letter[5] from Salam in response marked 'PERSONAL AND CONFIDENTIAL'. He wrote 'I am somewhat puzzled at the references'. We had referred to the Weinberg model in our paper. Clearly Salam would have preferred us to have referred to the 'Weinberg-Salam model' which was a term increasingly used at the time. He went on 'Since I may claim that we know each other rather well, I wish you had asked me to comment before on what you were proposing to write'. I responded that in his 1968 paper he had not done everything that Weinberg had done in 1967: more specifically he had not established the relationship between the masses of the W- and Z-mesons and therefore would have been unable to predict a

"particular neutral current rate given the corresponding charged current rate"[6].

So did Salam do what the New York Times (presumably based on a briefing by the Nobel Committee) had claimed that he had done; namely propose a theory in 1967 which predicted the size of the parity-violation in electron scattering off hydrogen nuclei as observed in 1979. If he didn't why was he awarded the Nobel prize? I knew all three prize winners and NATURE had asked me to write about the award of the prize in November 1979[7]. So I would like to explain how Salam won the prize even though he was unable to predict the result of the 1979 experiment.

Abdus Salam was born in 1926 in the Punjab, then in British India and now in Pakistan[8]. (He died in 1996). His family wanted him to aim for the Indian Civil Service and he won a scholarship to Government College Lahore aged 14 with the highest marks ever recorded. In 1946 he won a Punjab Government scholarship to St John's College, Cambridge, where he was assigned the well-known cosmologist Fred Hoyle as supervisor. Salam got a first in Part II of the Mathematics Tripos after two years and a first in Physics in his third year. He then began research under the supervision of Paul Matthews who had just completed his Ph. D. himself. At the time the big topic in theoretical physics was quantum electrodynamics or QED--the study of how electrons interact with photons (photons are the 'particles' or quanta of light required by quantum theory) and in particular how to remove the infinities from QED calculations through the process called renormalisation. Salam and Matthews jumped in and applied the techniques successfully applied by Richard Feynman, Julian Schwinger, Shin'ichirō Tomonaga and Freeman Dyson in QED to more general processes. For this Salam received the 1950 Smith's Prize, awarded to the Cambridge graduate student who has made the greatest progress in mathematics or theoretical physics. In 1951 he joined Matthews at the Institute of Advanced Study in Princeton whose Director was Robert Oppenheimer, the head of the wartime Manhattan Project to build an atomic bomb and where Dyson was a Fellow. Salam had arrived in the premier division of theoretical physics.

As a good patriot Salam then returned to Pakistan as Professor of Mathematics at both Punjab University and Government College. He hoped to continue his research but he found that he had little time and no encouragement from his superiors for research: for extra-curricula activities he was expected to look after the football team. Moreover he had no access to physics journals. He was a member of the Ahmaddiya sect of Shia Islam which was viewed by Pakistan's orthodox Sunni population as heretical. In 1953 there were widespread anti-Ahmaddiya riots and Salam was advised he might be a target so in 1954 he returned to Cambridge as Lecturer in Theoretical Physics and Fellow of St John's College to resume his collaboration with Paul Matthews.

The mid-1950s were exciting times for physicists. The results of the theory of quantum electrodynamics were tested experimentally and agreed with the new theoretical results to better than one part in a million while new particles were being discovered almost every month. There were even hints that fundamental symmetries of nature were being violated: the Chinese American physicists T. D.Lee and C. N. Yang had suggested in January 1957[9] that parity could be violated in nuclear beta-decays so that the reflection of a beta-decay in a mirror would not represent a physically-allowed process. Salam then discovered[10] that if a particle called a neutrino which had no electric charge and was emitted together with an electron in beta-decay had precisely zero mass (and thus according to Einstein's theory of relativity must travel at the velocity of light) then the equation which describes it has a natural symmetry which violates parity so that a world in which left-handed was preferred to right-handed could be the consequence of zero mass neutrinos. Paul Matthews wrote to him from the United States 'you've really hit the jackpot this time'[11]. His friend and collaborator John Ward wrote 'So many congratulations and fond hopes for at least one-third of a Nobel prize'[12]. One-third because the prize would presumably be shared with Lee and Yang.

Unfortunately for Salam the Nobel Committee didn't agree. After experimental confirmation[13] in 1957 that parity was violated and that neutrinos emitted in beta-decays rotated preferentially anti-clockwise (that is were left-handed) the 1957 Nobel prize was awarded to Lee and Yang not Salam. Salam continued to hope that future Nobel Committees would take a more positive view of his work.

By the mid-1950s the UK was recovering from the effects of the Second World War--it was the time of Harold Macmillan's 'you've never had it so good' and Patrick Blackett, Physics Nobel laureate and scientific advisor to successive governments was head of the Physics department at Imperial College London. He had had a good war in Operational Research, had excellent contacts with UK funding agencies and wished to build up a physics department equal to its rivals at MIT, Harvard and Princeton. He needed to appoint a young theoretical physicist who was in contact with the latest results on QED as professor to head the theory group. Having taken soundings from Hans Bethe, the leader of the theoretical physics group in the Manhattan Project, Blackett approached Salam who agreed to move to London provided that Matthews came with him. Salam came in January 1957 at the age of 31 and was elected a Fellow of the Royal Society two years later. He was then the youngest Fellow and the first Asian to hold a chair in a science faculty in the UK.

While Salam was at the Institute in Princeton in 1951, Zafrullah Khan, the Foreign Minister of Pakistan and a fellow Ahmadi whom Salam had previously met en route to Cambridge in 1946 visited him on his way to the UN General Assembly. They toured New England together and then Khan showed Salam round the UN in New York. So in 1955 at the first UN Conference on Peaceful Uses of Atomic Energy in Geneva, Salam found himself appointed as a scientific secretary to the conference. With backers like Zafrullah Khan, Blackett, Bethe and Oppenheimer, Salam had qualified for the premier league in international scientific politics as well as theoretical physics.

The main job of the conference was to discuss the setting up a new UN Agency which would safeguard fissionable material worldwide so states could benefit from nuclear power without leading to nuclear weapon proliferation. President Eisenhower had offered small reactors to developing countries in December 1954 as part of his Atoms for Peace programme. Following the 1955 conference, 81 nations unanimously approved the statute of a new International Atomic Energy Agency (IAEA) in October 1956. A second UN conference on Peaceful Uses of Atomic Energy was held in Geneva in 1958 devoted to civil nuclear power and in particular to the possibilities of using thermonuclear fusion to produce electricity.

Salam was a novice in all this. His education encompassed neither nuclear reactor theory nor International Relations. But he learned fast. He noticed in particular the leading role played by Swedes in UN business. Although the Indian physicist Homi Bhabha presided at the 1955 meeting, the Swedish nuclear physicist Sigvard Eklund was elected the Secretary General of the 1958 meeting while Dag Hammarskjold was the Secretary General of the UN itself. And a fellow scientific secretary at the 1955 conference was Ivar Waller, the veteran Swedish theoretical physicist who had made fundamental contributions to the understanding of crystal lattices and to the importance of negative energy states in the Dirac equation . Wallar and Salam became good friends as did Eklund and Salam. Eklund became the Director of the IAEA in 1961. Salam later wrote 'from that date [1958] started a most cherished personal friendship and one that transformed my life'[14]. Eklund was a nuclear physicist who worked on reactors and had been employed by the Nobel Institute in Stockholm for 8 years before moving on to a career in academia, government and industry. He knew everyone in the Swedish scientific and political establishment. Salam didn't remain a novice for very long and learned quickly about the scientific hierarchy in Sweden.

He had been forced to leave Pakistan for the UK in order to pursue his career in theoretical physics and wondered whether scientists from third world countries could be helped to continue their research fruitfully in their home country without having to move to developed countries. Salam conceived the idea of a centre of excellence to which scientists from the Third World could come on a regular basis for visits of a few weeks or months, to keep in touch with research at the frontier of their subject, but still remain for most of the year working in their own countries. Provided the subject matter was theoretical so expensive laboratory equipment wasn't required, provision of such a centre should not require expenditure of more than a few hundred thousand dollars to start up and a similar amount in running costs. These were tiny sums compared with the budgets for research of states like the US, UK and Soviet Union.

Salam was a devout Moslem throughout his life. He believed that miracles are possible but must be helped to happen. He considered that his journey from a small town in the Punjab to Cambridge illustrated this. A sum of 150,000 rupees had been collected by the governing Muslim Union in the Punjab to support the allied war effort. When the war ended the cash for no particular reason was assigned for a scholarship for poor farmer's sons to study abroad. Salam's uncle had just bequeathed a piece of land to Salam's father so Salam turned out to be eligible for the scholarship. Then St John's College Cambridge was expecting an Indian student who was unable to travel. Salam's hard work had translated into an outstanding academic record and this secured him the scholarship so that he could be admitted immediately for admission in October 1946. The scholarship wasn't renewed: Salam was the sole recipient. Thus Salam found himself at St John's College where Britain's pre-eminent theoretical physicist and Nobel laureate Paul Dirac was a Fellow: this sequence of events did indeed seem a miracle, albeit helped by his hard work.

In 1958 General Ayub Khan seized power in a coup.. He asked Salam to help establish the Pakistan Atomic Energy Commission and in 1961 appointed him Chief Scientific Advisor. Salam was also appointed as Pakistan's representative to the General Conference of the

---

[14] A Salam, *Ideals and Realities* (Ed. Z Hassan and C H Lai) World Scientific Singapore 1984 p.52

IAEA. Salam's old friend from his student days at Government College, nuclear engineer Munir Ahmad Khan, had recently been appointed to a senior staff position at the IAEA in Vienna and after taking advice from him Salam tabled a resolution at the General Conference that the IAEA should set up a study group to consider the establishment of an international centre for theoretical physics which would cater for the needs of physicists from developing countries. It was difficult for anyone to disagree with a study group so the resolution was passed: Salam and three close colleagues then went to work to write a proposal. Paolo Budinich, a physicist from Trieste, heard about this while in Rome and suggested that the Italian government recommend Trieste as a site for the proposed institute: Trieste at the time was just recovering from being under military occupation—allied forces had only left in 1955 –and both the Italian government and most of its inhabitants were desperate to cement Trieste's ties with Italy rather than with Yugoslavia and to obtain international recognition that Trieste was now Italian. An international centre sponsored by a UN agency with Soviet agreement was just the job. Italy offered both a site and co-sponsorship of the Centre with the IAEA. Although there was substantial opposition from developed countries and even IAEA's own Scientific Advisory Committee, Eklund and Munir Khan helped Salam prevail and the ICTP opened in 1964 with Salam as Director and Budinich as his deputy. UNESCO joined ICTP's sponsors shortly afterwards

Spurred by Ward's and Matthews' letters, Salam now could turn his attention to his prime goal. On the wall of his office in Trieste he put the Persian prayer 'O Lord, work a miracle!' And he set to work hard to make it happen using ICTP's resources: the miracle he wished for was the Nobel Prize in Physics. Dirac retired to Florida State University and received a standing invitation to visit ICTP with expenses paid[15]; Waller visited every summer on a similar basis[16]. Waller was on the Nobel Committee from 1945 until 1972. In summer 1972 Salam convened a Conference on the history and foundations of quantum mechanics at ICTP together with a conference banquet in honour of Dirac's 70[th] birthday. This was unrelated to the needs of developing countries but allowed Salam to fraternise with important physics dignitaries: past Nobel prize winners Dirac, Heisenberg, Wigner and Bethe all attended together with other leading theoretical theoretical physicists such as Casimir from Holland and Peierls from Oxford.

Salam's own research in elementary particle theory naturally became the focus of the work at the Centre in its early days although it had almost nothing to do with IAEA's mission of enlarging the contribution of atomic energy to the world. Moreover a centre of theoretical physics devoted to the needs of developing countries could be expected to concentrate on areas such as solid state physics, which deals with the semi- conductors used in electronic devices, rather than with particle physics which is of no direct use for anything. And yet every time I went to visit ICTP in its first ten years of operation it seemed to me to be just the particle theory group at Imperial transposed 600 miles east to the seaside and warmth. Following advice from Salam's Scientific Council that the programme at ICTP was unbalanced and needed to be diversified away from particle physics, Salam asked his old Cambridge friend John Ziman to start off a condensed matter programme in 1967. This was successful and by 1970 Salam needed to find a Director of the programme who would run an annual three-month course every summer, together with an extended course every second year. Salam consulted Ziman and Waller and chose Stig Lundqvist of Chalmers University of Technology in Gothenburg, a former student of Waller. According to a Scandinavian physicist who has known members of the Nobel Committee for many years "He [Salam] must have realized in the 60's that the Nobel Committee would need a modern condensed matter person and since Lundqvist was Waller's student he put his money on him….. I remember seeing Abdus several times here... [before 1971] and that he flirted with Stig Lundqvist then."[17]

Salam's bet came off: Lundqvist became a member of the Nobel Committee in 1973 and stayed for twelve years. Moreover according to Lundqvist at Salam's memorial conference 'I had very close contact with Abdus Salam. We discussed the scientific programme and above all the interesting new physics he was doing. The possibility of a Nobel prize was coming close and as I was a member of the Nobel Committee, these discussions became very complex"[18]. I bet they did. But not complex enough for hard questions about who did what. Lundqvist after all was not a particle physicist and he would have fallen, like the New York Times, for the story of the success of the Weinberg-Salam model.

So what is the real story of the Weinberg-Salam model? Glashow[19] wrote a paper in 1961 which incorporated the electromagnetic current with the charged weak current observed in nuclear beta-decay in the same theory. [In beta-decay the neutrino is always accompanied by a charged electron or positron to constitute a charged current]. In so doing Glashow predicted that neutrinos and electrons could scatter off other particles without exchange of charge through a neutral weak current. The form of the neutral weak current was given by the

theory—in particular it violated parity—but not its strength. In 1967 Weinberg[20] applied the Higgs mechanism [named after Peter Higgs of Edinburgh University] to the Glashow theory thereby predicting the mass of the Z-meson and so giving the strength of the neutral weak current. This allowed the prediction of the parity-violation in polarized electron scattering observed at Stanford twelve years later. Furthermore in principle the mechanism allowed the theory to be renormalisable; that is to say any calculation in the theory would be finite. This theory is what I and others called the Weinberg model.

Salam was perpetually sorry that his idea in a draft paper in 1956 suggesting that parity might be violated had not been published after criticism from senior colleagues. His approach to physics afterwards changed from rigorous to scattergun: he wrote (usually with collaborators) about a paper every month[21]. It was always possible that one might turn out to be right. Even in his letter rebuking me for not using the term Weinberg-Salam, he enclosed a new preprint describing another alternative theory which, like Bailin and mine, sank without trace. Nor did he worry that other people may have had the same idea. In 1964 he and Ward[22] wrote a paper on electromagnetic and weak interactions which should not have been accepted for publication because it just repeated Glashow's work of 1961.

Salam gave a graduate course at Imperial College most years and in autumn 1967 the course focused on the theory of electromagnetic and weak interactions and the possible application of the Higgs mechanism to the theory. No written record of the course exists although at least one participant remembers it[23]. Salam did not submit an article on the application of the Higgs mechanism to electroweak theory for publication in a peer-reviewed journal. But around that time he was asked to be a member of the Advisory Committee for a Symposium to be held in Gothenberg in May 1968 on particle theory, sponsored by the Nobel Foundation. By 1968 there was no way that Salam could publish his lecture in a recognised peer-reviewed journal since Weinberg's paper had already been published, so Salam took the opportunity to give a lecture at the Symposium which claimed to be his lecture at Imperial the previous autumn. The proceedings of the Nobel Symposium were published as an expensive monograph[24] with circulation limited to a few specialist libraries. Hardly any of the more than 1500 of physicists who have cited Salam 1968 in their papers have read the paper. It is still difficult to find although a copy[25] is now available on the internet. As is clear from Salam's letter to me, any deviation from the name Weinberg-Salam would bring a rebuke from Salam, and since most physicists hadn't read the 1968 paper, the name Salam-Weinberg won general acceptance especially after Ben Lee apologised to Salam following such a rebuke[26]. Waller and four other current or future members of the Nobel Committee were present at the Symposium[27] to hear Salam's version of who did what. Glashow does not get a mention. More to the point, Murray Gell-Mann who gave the concluding talk (and who knew exactly who did what) doesn't refer to Salam's contribution in his Summary of the Symposium[28]!

Suppose Salam did give a lecture in autumn 1967 saying what he then published in the Nobel Symposium. It still was not the same as Weinberg's paper because the relative strength of the neutral and charged weak currents had not been calculated. So a calculation based on Salam's paper could not have predicted the result of the experiment at Stanford in 1979. Nor did Salam himself consider that his Gothenburg lecture broke new ground. On 1[st] October 1969 he wrote to Ivar Waller who then was the most senior member of the Nobel Committee 'In accordance with your wish, I am setting down my contributions to neutrino and weak interaction physics'[29]. Salam then described his 1957 paper showing that a zero mass neutrino would naturally lead to a theory of beta-decay in which parity was violated. This was a nice result but not exceptional; the Soviet physicist Lev Landau[30] published a more general formulation within two months of Salam, and Feynman and Gell-Mann[31] had worked out a much more detailed theory of weak interactions within a year. The letter to Waller contained four and a half pages. It focuses almost exclusively on his 1957 paper and its extension by others. He only has two lines devoted to what is now called the Weinberg-Salam model namely 'With Ward I was the author in 1963 of a gauge theory of weak interactions-about which I spoke at the Nobel Symposium in Gothenburg in 1968. [The paper[22] with Ward was the paper which repeated Glashow's work]. The chief characteristic of this is the natural appearance of neutral currents, which as you are aware, are now stirring again' [That is true but it was Glashow[19] who first predicted neutral currents]. So less than 18 months after the Gothenburg lecture, Salam hardly mentions it in his privileged letter to the senior member of the Nobel Committee in which he wrote down his most important physics contributions.

From 1971 onwards Dirac nominated Salam for the Prize[32].   Salam wrote a supporting letter on his own behalf and passed it to Paul Matthews to send on to the Nobel Committee[33]. The index to Salam's papers  reads  "A81 Manuscript draft letter to P.T.Matthews composed by Salam in support of his own possible Nobel candidacy [1971]. Matthews dutifully  sent it off on his own notepaper. Each year a new Matthews letter drafted by Salam was sent  to a senior member  of the Nobel Committee.  Waller left the Committee in 1972 so the letter was modified and addressed to Bengt Edlen[34].  Edlen left   in 1976 when Hulthen became Chairman. Salam then instructs Matthews 'I think the same letter could go to Hulthen and I now feel you should send it. In your letter you can enclose …my letter to Waller'. Salam concludes 'So let us trust to God and send the stuff'[35].

Then there was the awkward matter of the date of Salam's contribution to the Gothenburg Symposium being 1968 whereas Weinberg published in 1967. Matthews to the rescue again[36]: the index reads "Copy of a letter in which P T Matthews confirms that he heard Salam describe the unified gauge theory of weak interactions in a lecture in 1967 written in support of Salam's Nobel Prize candidacy". But Matthews' letter gives no detail of what was in that lecture. So the prize was awarded on the basis of a non-peer-reviewed publication which quotes an unpublished lecture. Yet  the name Weinberg-Salam model stuck because almost everyone used it. Weinberg was  happy with the name Weinberg-Salam:  he knew that he could only benefit from association with Salam. Glashow in fact was nearly excluded  but Gell-Mann informed  the Nobel Committee of his contribution just in time[37].

Salam's prayers were answered on October  15 1979.

---

ACKNOWLEDGEMENTS

I would like to thank David Bailin, Shelly Glashow, Tom Kibble and Jogesh Pati for their help. I am grateful to Frank Close for early sight[38] of some material from his book[39] which presents an independent summary of Salam's role in the 1979 Nobel prize: in particular the memos from Salam to Matthews[33] in which Salam writes his own nomination letter. I would also like to thank the ICTP staff for their cooperation when I visited in March 2011.

---

[38] In November 2010 Close sent me some extracts from his book outlining his conclusions and which describe the origins of the Weinberg-Salam model./
[39] Frank Close The Infinity Puzzle, OUP (UK); Basic Books (US), October 2011